\documentclass[structabstract]{AA}
\usepackage{graphicx}             
\usepackage{txfonts}
\begin{document}
\renewcommand{\textfraction}{0.0}
\renewcommand{\bottomfraction}{1.0}
\renewcommand{\dblfloatpagefraction}{1.00}

   \title{Star formation associated with a large-scale infrared bubble}
  \author{Jin-Long Xu
          \inst{1,2}
          and Bing-Gang Ju\inst{3,4}
          }
   \institute{ National Astronomical Observatories, Chinese Academy of Sciences,
             Beijing 100012, China \\
         \email{xujl@bao.ac.cn}
          \and
          NAOC-TU Joint Center for Astrophysics, Lhasa 850000, China
         \and
          Purple Mountain Observatory, Qinghai Station, 817000, Delingha, China
          \and Key Laboratory of Radio Astronomy, Chinese Academy of Sciences, Nanjing 210008, China
        }
\authorrunning{J.-L. Xu and B.-G. Ju}
\titlerunning{Star formation associated with a large-scale infrared  bubble}
   \abstract
   {}
{To investigate how a large-scale infrared bubble centered at $l$=53.9$^{\circ}$ and $b$=0.2$^{\circ}$ form, and study if star formation is taking place at the periphery of the bubble, we performed a multi-wavelength study.}
{Using the data from the Galactic Ring Survey (GRS)  and Galactic Legacy Infrared Mid-Plane Survey Extraordinaire (GLIMPSE), we performed a study for a large-scale infrared  bubble with a size of about 16 pc at a distance of 2.0 kpc. We present the $^{12}$CO $J$=1-0, $^{13}$CO $J$=1-0 and C$^{18}$O $J$=1-0 observations of HII region G53.54-0.01 (Sh2-82) obtained at the the Purple Mountain Observation (PMO) 13.7 m radio telescope  to investigate the detailed distribution of associated molecular material. In addition, we also used radiorecombination line and VLA data. To select young stellar objects (YSOs) consistent with this region, we used the GLIMPSE I catalog.  }
{The large-scale infrared bubble shows a half-shell morphology at 8 $\mu$m. H {\footnotesize II} regions G53.54-0.01, G53.64+0.24, and G54.09-0.06 are situated on the bubble. Comparing the radio recombination line velocities and associated $^{13}$CO $J$=1-0 components of the three H {\footnotesize II} regions, we found that the 8 $\mu$m emission associated with H {\small II} region G53.54-0.01 should belong to the foreground emission, and only overlap with the large-scale infrared bubble in the line of sight. Three extended green objects (EGOs, the candidate massive young
stellar objects), as well as  three H {\small II} regions and two small-scale bubbles are found located in the G54.09-0.06 complex, indicating an active massive star-forming region. C$^{18}$O $J$=1-0 emission presents four cloud clumps on the northeastern border of  H {\small II} region G53.54-0.01. Via comparing the spectral profiles of $^{12}$CO $J$=1-0, $^{13}$CO $J$=1-0, and C$^{18}$O $J$=1-0 peak at each clump, we found the collected  gas in the three clumps, except for the clump coincided with a massive YSO (IRAS 19282+1814). Using the evolutive model of H II region, we derived that the age of H {\footnotesize II} region G53.54-0.01 is 1.5$\times10^{6}$ yr. The significant enhancement of several Class I and Class II YSOs around G53.54-0.01 indicates the presence of some recently formed stars, which may be triggered by this H {\footnotesize II} region through the collect and collapse (CC) process.
 }
   {}


   \maketitle
%

\section{Introduction}
Star formation requires dense self-gravitating gas. The activity of ultraviolet (UV) radiation, stellar winds, and  supernova
remnant (SNR) can all compress or accumulate a pre-existing gas into a dense gas. The gas may become gravitational instabilities, then collapse into dense cores. Such three dynamic processes may trigger the formation of a new generation of stars. Churchwell et al. (\cite{Churchwell06,Churchwell07}) compiled a list
of $\sim$600 objects with 8 $\mu$m emission in a  bubble morphology, suggesting that the bubbles were polycyclic aromatic hydrocarbon
(PAH) emission in the photodissociation regions (PDRs) surrounding
O and early-B stars. Deharveng et al. (\cite{Deharveng10}) studied 102 bubbles catalogued by Churchwell et al. (\cite{Churchwell06,Churchwell07}),  most of the bubbles enclose H {\small II} regions ionized by O-B2 stars. They also concluded that the bubbles may trigger 14$\%$$\thicksim$30$\%$ of the star formation in our Galaxy. Recently several pieces of observational evidence have been found in favor of star formation triggered by bubbles (see e.g. Dewangan et al. \cite{Dewangan12}; Ji et al. \cite{Ji12}; Zhang et al. \cite{Zhang13}; Dewangan \& Ojha \cite{Dewangan13}; Li et al. \cite{Li14}).

Leahy et al (\cite{Leahy08}) identified a large-scale infrared  bubble at 8 $\mu$m from the Galactic Legacy Infrared Mid-Plane Survey Extraordinaire (GLIMPSE) project, which just surrounds a large radio shell (centered at $l$=53.9$^{\circ}$ and $b$=0.2$^{\circ}$) with an angular size of 30$^{\prime}$$\times$26$^{\prime}$. Velusamy et al. (\cite{Velusamy86}) classified the radio shell as an H {\small II} region due to possible thermal infrared emission. With a higher spatial resolution, Leahy et al (\cite{Leahy08}) showed that the radio shell and infrared bubble are physically separated, suggesting that the radio shell may be an old SNR at a distance of $\simeq$7.3 kpc. There are three H {\small II} regions (G53.54-0.01, G53.64+0.24, and G54.09-0.06) and a young SNR (G54.1+0.3) in the infrared  bubble,  as seen in Fig. 1.

H {\small II} region G53.54-0.01(Sh2-82) coincides with the bubble N115 (Churchwell et al. \cite{Churchwell06}; Deharveng et al. \cite{Deharveng10}; Sherman \cite{Sherman}). HD 231616 with a B0V type and a mass of 18 M$_{\odot}$ may be the exciting star of G53.54-0.01 (Hunter \& Massey \cite{Hunter90}). Using the Galactic Ring
Survey (GRS) data, two clumps of $^{13}$CO $J$=1-0 were identified at the northeast of G53.54-0.01 (Yu et al. \cite{Yu}). G54.1+0.3 is a core-collapse SNR with centrally brightened synchrotron emission in radio and X-rays (Velusamy \& Becker \cite{Velusamy88}; Lu et al. 2002), which closely resembles the Crab Nebula. HI and CO observations give the kinematic distance of $\simeq$ 6.2 kpc to G54.1+0.3 (Leahy et al. \cite{Leahy08}).  The SNR is associated with a pulsar (PSR J1930+1852), which has a 136 ms rotational period and a characteristic age of 2900 yr (Camilo et al. \cite{Camilo02}; Lu et al. \cite{Lu08}). Koo et al. (\cite{Koo08}) detected a star-forming loop around SNR G54.1+0.3 using the AKARI infrared satellite. They proposed that the star-forming loop is triggered by the progenitor star of
G54.1+0.3.

\begin{figure*}[]
\vspace{0mm}
\centering
\includegraphics[angle=0,scale=0.54]{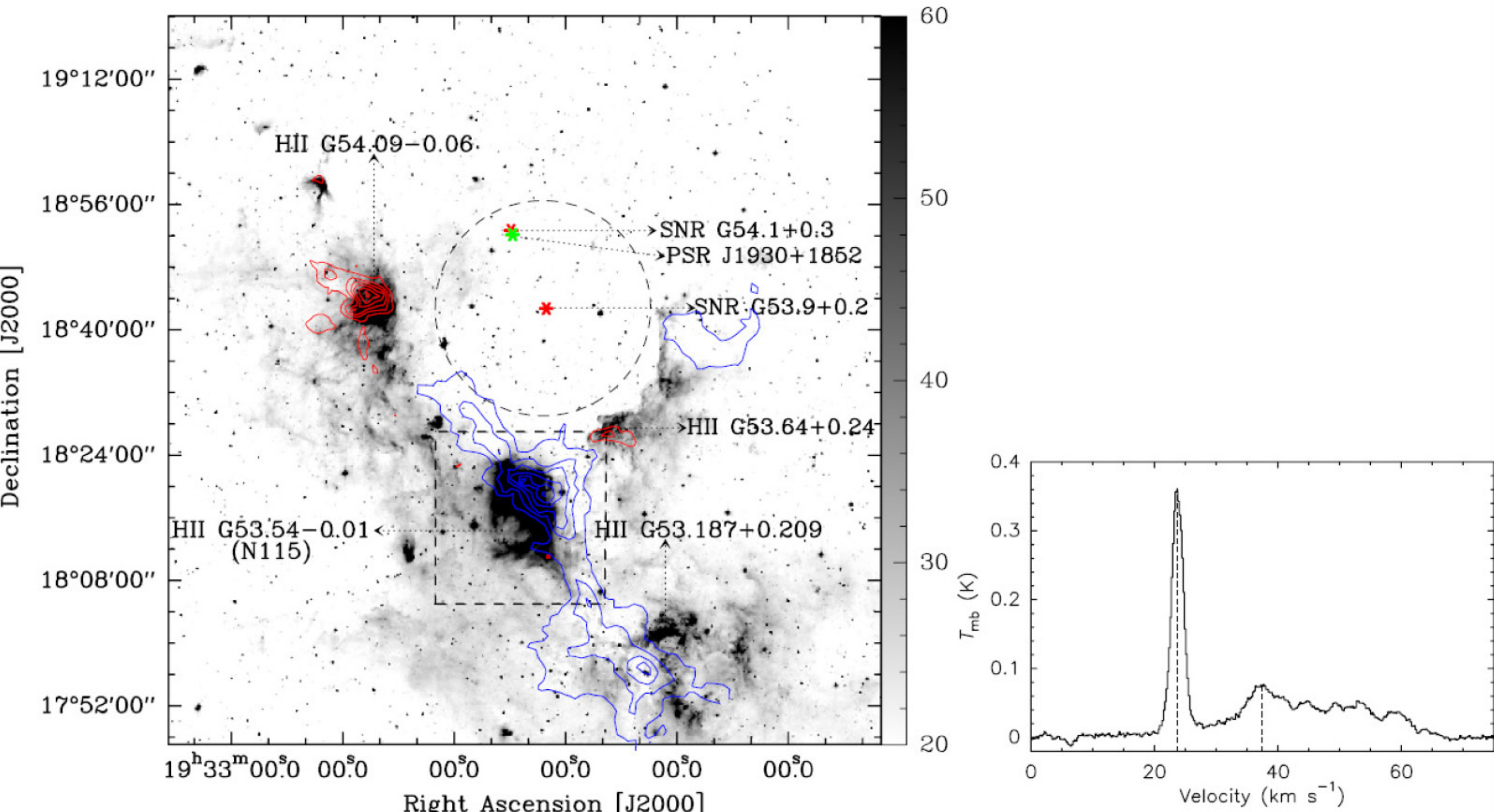}
\vspace{-2mm}\caption{Left panel: $^{13}$CO $J$=1-0 integrated intensity
contours (blue and red) overlayed on the Spitzer-IRAC 8
$\mu$m emission map (grey scale). The blue contour
levels are 20, 35,... , $95\%$ of the peak value (19.9 K km
s$^{-1}$) integrated from 20 to 28 km
s$^{-1}$, while the red contour
levels are 20, 35,... , $95\%$ of the peak value (26.1 K km
s$^{-1}$) integrated from 35 to 43 km
s$^{-1}$.  The different sources associated with the region are marked. The red and green stars represent the center of SNRs and PSR. The radius of the radio shell is 14$^{\prime}$ outlined by a dashed circle. The dashed quadrate frame marks the area observed with PMO telescope.  The unit of the grey bar is in MJy sr$^{-1}$. Right panel: Average spectra of $^{13}$CO $J$=1-0 over the whole large-scale infrared bubble. The vertical dashed lines in the spectra mark the peak velocity.}
\end{figure*}

To explore signatures of star formation in a large-scale infrared bubble, we mainly combined
molecular, infrared, and radio continuum
observations toward H {\small II} regions G53.54-0.01 and G54.09-0.06. The observations and data reduction are described in
$\S$2, and the results are presented in $\S$3. In $\S$4, we discuss
how our data provide evidence of triggered star formation in the large-scale infrared
bubble. The conclusions are summarized in $\S$5.

\section{Observations and data reduction}
\subsection{$^{13}$CO $J$=1-0 GRS data}
Galactic Ring Survey (GRS) is a survey of $^{13}$CO $J$=1-0 emission (Jackson et al. \cite{Jackson06}). The survey covers a longitude range of $\ell$$=$18$^{\circ}$--55.7$^{\circ}$ and a latitude range of $|b|$$\leq$1$^{\circ}$, with a  angular resolution of 46$^{\prime\prime}$. The survey's velocity coverage is -5 to 135 km s$^{-1}$ for Galactic
longitudes $\ell$$\leq$40$^{\circ}$ and -5 to 85 km s$^{-1}$ for Galactic longitudes $\ell$$>$40$^{\circ}$. At the velocity resolution of
0.21 km s$^{-1}$, the typical rms sensitivity is 0.13 K. We used the GRS archival data{\bf \footnote{http://www.bu.edu/galacticring/}} to study the molecular emission of the large-scale infrared  bubble.

\begin{figure*}[]
\vspace{0mm}
\includegraphics[angle=0,scale=1.23]{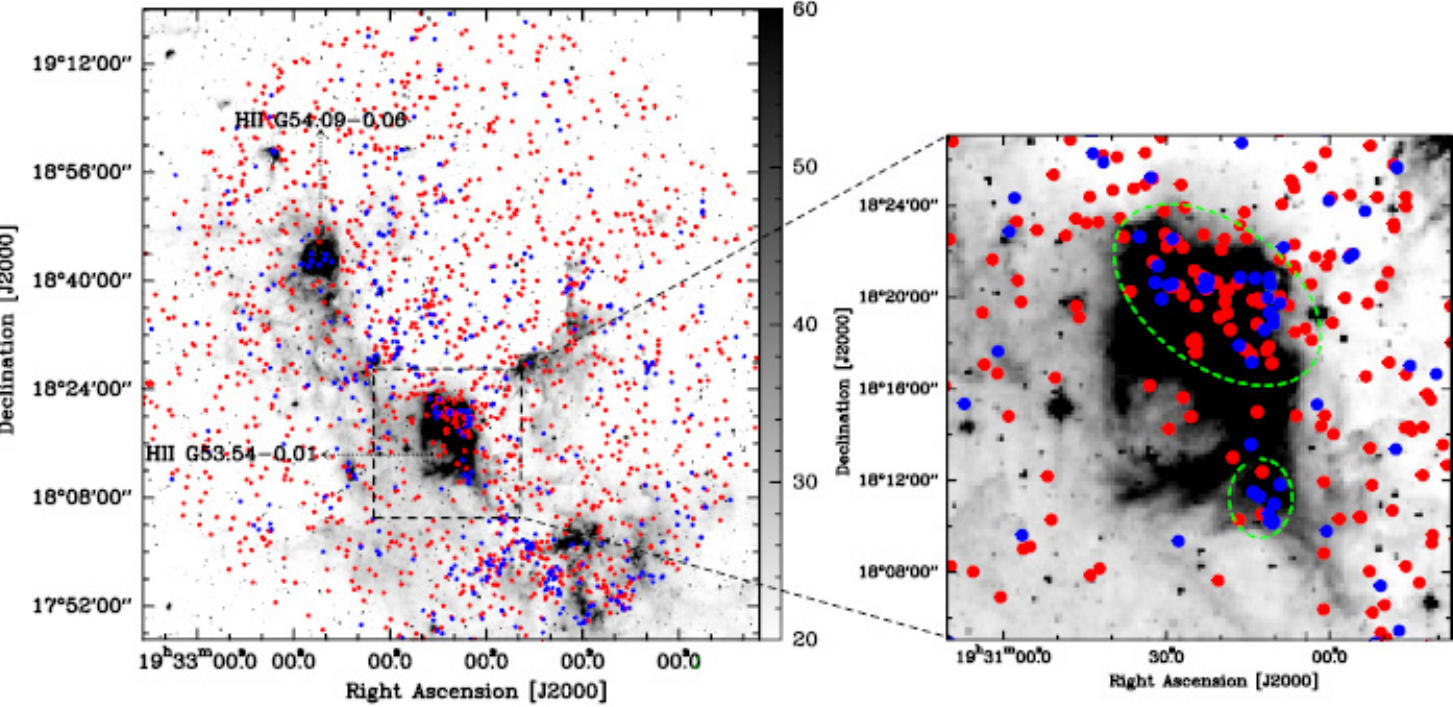}
\vspace{-2mm}\caption{panels: The Spitzer-IRAC 8
$\mu$m emission map (grey scale) overlayed on Class I and Class II sources labeled as the
blue and red dots, respectively.  The unit of the grey bar is in MJy sr$^{-1}$. Right panels: The distributed map of Class I and Class II sources in the vicinity of H {\small II} region G53.54-0.01. Two green dashed ellipses mark the concentration of Class I and Class II sources. }
\end{figure*}

\subsection{Purple Mountain data}
The mapping observations of H {\small II} region G53.54-0.01 and its adjacent region were performed in the $^{12}$CO
$J$=1-0, $^{13}$CO $J$=1-0, and C$^{18}$O $J$=1-0 lines using the Purple Mountain Observation (PMO) 13.7 m radio telescope (Zuo et al. \cite{Zuo04}) at De Ling Ha in the west of China at an altitude of 3200 meters, in May  2013.
The Superconducting Spectroscopic Array Receiver (SSAR) is used, which is a superconducting focal
plane array that observes a 3$\times$3 beam.
The new array receiver system in single-sideband (SSB) mode was used as front end. The back end is a fast Fourier transform spectrometer (FFTS) of 16384 channels with a bandwidth of 1 GHz,
corresponding to a velocity resolution of 0.16 km s$^{-1}$ for $^{12}$CO
$J$=1-0 and 0.17 km s$^{-1}$
for $^{13}$CO $J$=1-0 and C$^{18}$O $J$=1-0; $^{12}$CO
$J$=1-0 was observed at upper
sideband, while $^{13}$CO $J$=1-0 and C$^{18}$O $J$=1-0 were observed simultaneously
at lower sideband.  The half-power beam width (HPBW)
was 53$^{\prime\prime}$ at 115 GHz
and the main beam efficiency was 0.5. The pointing
accuracy of the telescope was better than 5$^{\prime\prime}$,  which was derived from continuum observations of planets.  W51D was observed once per hour as flux calibrator. The system noise
temperature (Tsys) in SSB mode varied between 150 K and 400
K. Mapping observations were centered at
RA(J2000)=$19^{\rm h}30^{\rm m}23.98^{\rm s}$,
DEC(J2000)=$18^{\circ}15'03.3^{\prime\prime}$ using on-the-fly (OTF) observing mode. The total mapping area is $15^{\prime}\times 15^{\prime}$ in
$^{12}$CO $J$=1-0, $^{13}$CO $J$=1-0, and C$^{18}$O $J$=1-0 with a $0.5^{\prime}\times0.5^{\prime}$
grid. The standard chopper wheel calibration technique was used to measure antenna temperature $T_{\rm A}$$^{\ast}$ corrected for atmospheric absorption. The final data was recorded in brightness temperature scale of $T_{\rm mb}$ (K). The data were reduced using the GILDAS/CLASS\footnote{http://www.iram.fr/IRAMFR/GILDAS/} package.

\subsection{Additional data }
The 1.4 GHz radio continuum emission data were obtained from the
NRAO VLA Sky Survey  (NVSS; Condon et al. \cite{Condon98}).
NVSS is a 1.4 GHz continuum survey
covering the entire sky north of -40$^{\circ}$ declination (\cite{Condon98})  with a noise of about 0.45 mJy/beam and the resolution of 45$^{\prime\prime}$.

The radio-recombination lines data were derived from the GBT survey (Anderson et al. \cite{Anderson11}). The survey have detected 448 previously unknown Galactic H {\small II} regions at X-band (9 GHz, 3 cm) in the Galactic zone 343$^{\circ}$$\leq$$\ell$$\leq$67$^{\circ}$ and $|b|$$\leq$1$^{\circ}$. The FWHM beam size
of the telescope is approximately 82$^{\prime\prime}$ at this frequency.

GLIMPSE survey is used to identify the young stars along H {\small II} regions, which observed the Galactic plane (65$^{\circ}$
$<$ $|l|$ $<$ 10$^{\circ}$ for $|b|$ $<$ 1$^{\circ}$) with the four
IR bands (3.6, 4.5, 5.8, and 8.0 $\mu$m) of the Infrared Array
Camera (IRAC) (\cite{Fazio}) on the Spitzer Space
Telescope. The resolution is from 1.5$^{\prime\prime}$ (3.6 $\mu$m) to
1.9$^{\prime\prime}$ (8.0 $\mu$m).

\section{Results}
\subsection{The large-scale infrared bubble}
\subsubsection{Infrared and $^{13}$CO $J$=1-0 emission}
Figure 1, left panel shows the $^{13}$CO $J$=1-0 integrated intensity map superimposed on the Spitzer-IRAC emission at 8 $\mu$m (grey scale).
The Spitzer-IRAC 8 $\mu$m emission presents a half-shell morphology corresponding mainly to
fluorescence from PAH molecules. The density of the inner edge is more dense in the large-scale infrared bubble. According to Leahy et al. (\cite{Leahy08}), the infrared bubble around a large radio shell with an angular size of 30$^{\prime}$$\times$26$^{\prime}$ contains three H {\small II} regions. Two extended H {\small II} regions G54.09-0.06 and G53.64+0.24 have a recombination line velocities of 42.1$\pm$0.8 km s$^{-1}$ and 38$\pm$1.8 km s$^{-1}$, respectively. The recombination line velocity of H {\small II} region G53.54-0.01 is 23.9$\pm$0.3 km s$^{-1}$ (Anderson et al. \cite{Anderson11}).

The CO maps contain velocity information that allows to disentangle
different molecular components along the line of sight. The right panel of Fig. 2 shows the average spectra of $^{13}$CO $J$=1-0 over the whole large-scale infrared bubble. From the spectra, we can see that there is a strong line that located
in the velocity intervals 20 to 28 km s$^{-1}$ peak at $\sim$24 km s$^{-1}$. Other velocity components overlap between 30 and 63 km s$^{-1}$. In the overlapping range of velocity, there is a relatively strong line located
in the velocity intervals 35 to 43 km s$^{-1}$ peak at $\sim$38 km s$^{-1}$. Using the velocity ranges of 20 to 28 km s$^{-1}$ and 35 to 43 km s$^{-1}$, we made the integrated intensity map of $^{13}$CO $J$=1-0 (Fig. 1,  right panel), marked by red and blue contours, respectively. In Fig. 1, the velocity component of 20 to 28 km s$^{-1}$ presents a filamentary structure, which  is associated  well with  the infrared emission of H {\small II} region G53.54-0.01. However, the filamentary structure extend toward the center of the large-scale infrared bubble.  A weak component of 20 to 28 km s$^{-1}$  is located on the western border of the infrared bubble. Additionally, the velocity  components of 35 to 43 km s$^{-1}$ are consistent with the infrared emission of H {\small II} regions G54.09.0.06 and G53.64+0.24. The $^{13}$CO $J$=1-0 emission related to H {\small II} region G54.09.0.06 shows a comet-like morphology. Hence, the large-scale infrared bubble may consist of two components, while the velocity component in interval 20 to 28 km s$^{-1}$ should belong to the foreground emission, and only overlap with other component in the line of sight.

\subsubsection{Search for young stellar objects}
To study star formation in the vicinity of the large-scale infrared bubble by detecting all the young stellar objects (YSOs) around the infrared bubble and looking at their position with respect to the ionized gas and molecular condensations, we used the Spitzer-GLIMPSE I catalog. Considering only sources that have been detected in the four Spitzer-IRAC bands, we found 153286 near-infrared sources centered on the infrared shell within a circle of 48$^{\prime}$ in radius. Allen et al. (\cite{Allen}) showed that YSOs have specific infrared colors depending on their masses and their evolutionary stages. Based on the color selection criteria of YSOs (Allen et al. \cite{Allen}), we found 478 Class I sources and 1423 Class II sources. Class I sources are protostars with circumstellar envelopes with an age of $\sim$10$^{5}$ yr, while Class II sources are
disk-dominated objects with  an age of $\sim$10$^{6}$ yr. The left panel of Fig. 2 presents the spatial distribution of Class I and Class II sources. From the left panel, we see that Class I and Class II sources are symmetric distributed across the whole selected region, but concentrated around  G53.54-0.01. In order to clearly show the distribution of YSOs  in the vicinity of H {\small II} region G53.54-0.01, we made a small-scale distributed map of the YSOs shown in the right panel of Fig. 2, which obviously presents the concentration of YSOs at the border of G53.54-0.01. We do not know if all the YSOs seen in the direction of G53.54-0.01 lie at the same distance with the H {\small II} region and associated with it. However, the concentration of YSOs in the surroundings of G53.54-0.01 indicates that the association is highly probable.

\begin{figure}[]
\vspace{0mm}
\centering
\includegraphics[angle=0,scale=0.39]{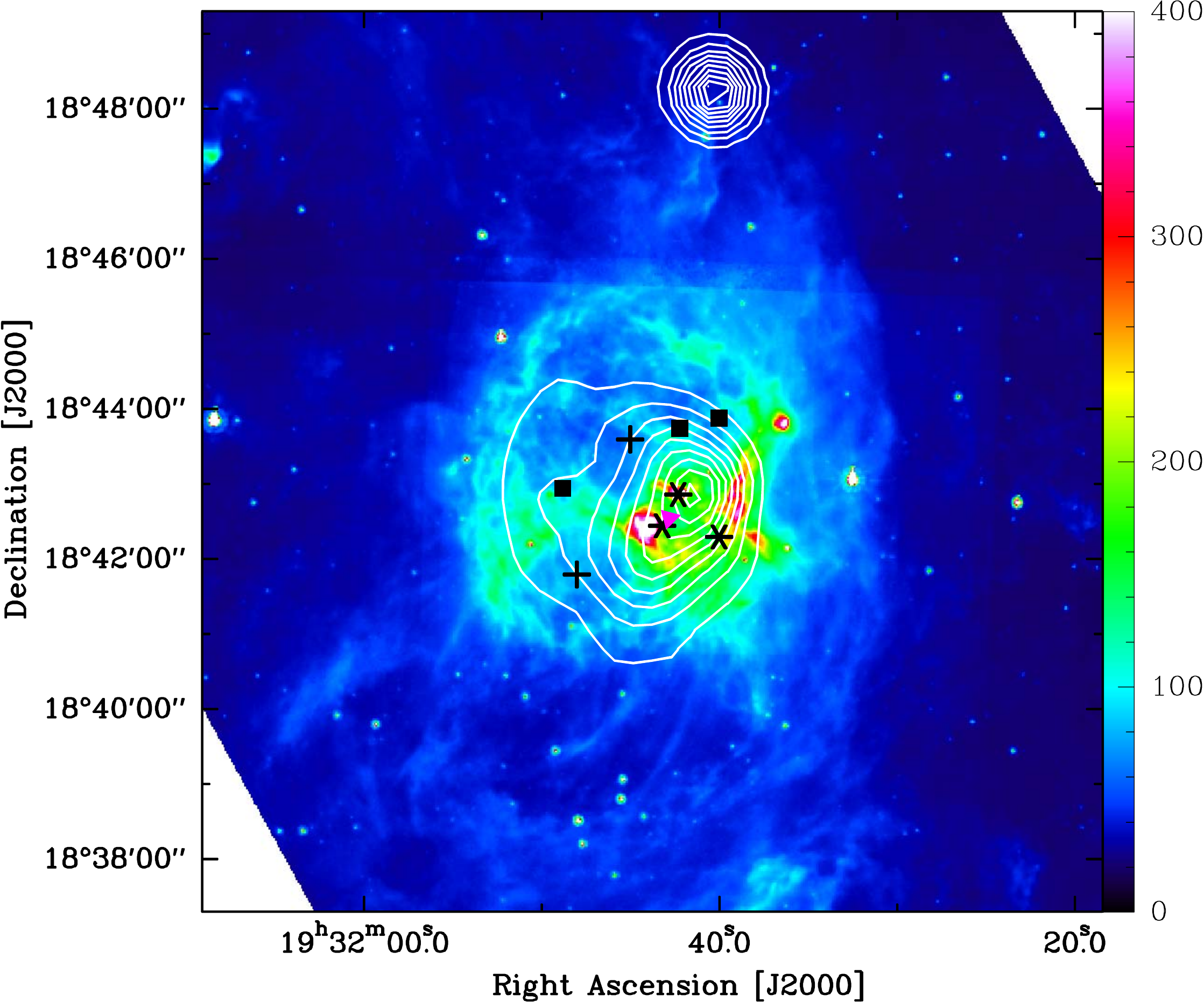}
\vspace{-1mm}\caption{1.4 GHz radio continuum emission
contours (white) overlayed on the Spitzer-IRAC 8
$\mu$m emission map (color scale). The  white contour levels are 20, 35,... , $95\%$ of the peak value (0.3 Jy beam$^{-1}$). Two black crosses indicate N116 and N117. $``\ast"$ show three HII regions. The EGO sources are labeled as the filled square symbols. The pink triangle represents the IRAS 19294+1836 source.  The unit of the color bar is in MJy sr$^{-1}$.}
\end{figure}

\subsection{H {\small II} region G54.09-0.06}
In Fig. 3,  the 1.4 GHz continuum emission obtained from NVSS  exhibits an extended
structure, which is consistent with the Spitzer-IRAC 8
$\mu$m emission.  From the survey of H {\small II regions at the radio band in the northern sky (Lockman \cite{Lockman89}), we found other two H {\small II} regions (G50.10-0.06 and G54.1-0.1), which may be associated with the Spitzer-IRAC 8
$\mu$m emission}. However, H {\small II} region G50.10-0.06 is closer to the center of the 1.4 GHz continuum emission. IRAS 19294+1836 with a velocity of 39.9 km s$^{-1}$ is an YSO, which is more near at the position of G54.09-0.06. Additionally, there are three extended green objects (EGOs) (G54.11-0.04, 54.11-0.05 and 54.11-0.08) in the region. EGOs have excess emission in extended structures in the Infrared Array Camera
(IRAC) 4.5 $\mu$m band images that are conventionally coded as
green in the IRAC false color images, which are considered as the candidate birth places of the massive YSO (Cyganowski et al. 2008). He et al. (2012) gave that the velocity of EGO 54.11-0.08 is 38.4 km s$^{-1}$, while that of EGO G54.11-0.04 is 39.2 km s$^{-1}$ (Chen et al. \cite{Chen10}), suggesting that this region is ongoing formation of many massive stars. Moreover, two  small-scale bubbles (N116 and N117) are located in this region, selected from the catalog of  Churchwell et al. (\cite{Churchwell06}) bubbles. Watson et al. (\cite{Watson10}) gave that the velocity of N117 is 18.7 km s$^{-1}$ from the survey of H {\small II} regions in the northern sky (Lockman \cite{Lockman89}), but we did not find the matching source with N117 in the survey results. As shown in Fig. 1 (right panel), the comet-like component of $^{13}$CO $J=1-0$  peak at 38 km s$^{-1}$  presents a good morphological correlation with the infrared emission containing G54.09.0.06. From Fig. 3, we also see that N116 and N117 are associated well with the Spitzer-IRAC 8 $\mu$m emission, suggesting that the velocity of  N116 and N117 may be $\sim$38 km s$^{-1}$.

\begin{figure*}[]
\vspace{0mm}
\includegraphics[angle=0,scale=.80]{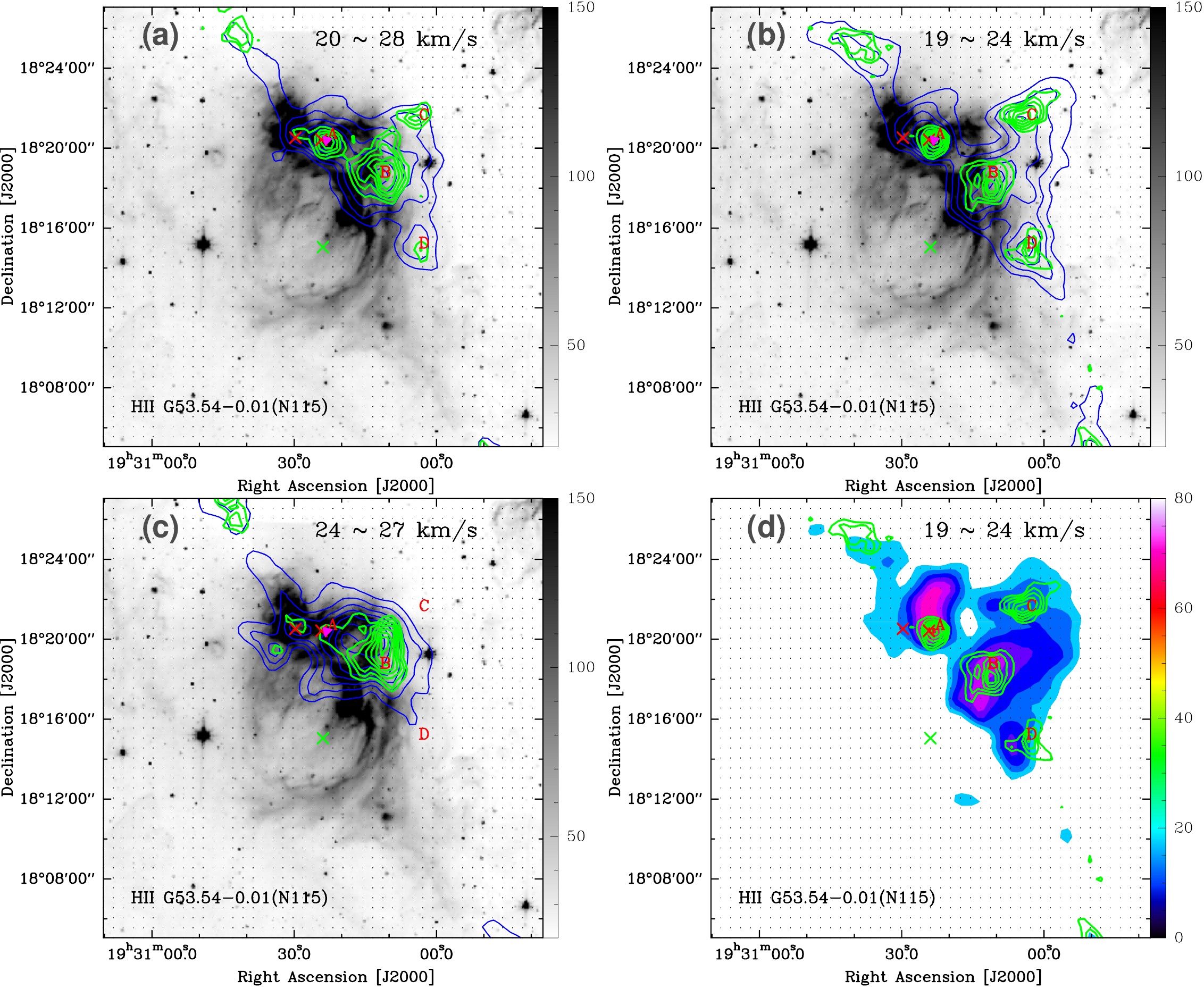}
\vspace{-5mm}\caption{ (a) The Spitzer-IRAC 8
$\mu$m emission map (grey scale) superimposed on $^{13}$CO $J$=1-0 (blue contours) and C$^{18}$O $J$=1-0 (green contours) intensity maps. The integrated velocity range is from  20 to 28 km
s$^{-1}$. (b) The integrated velocity range is from  19 to 24 km
s$^{-1}$. (c) The integrated velocity range is from  24 to 27 km
s$^{-1}$. (d) C$^{18}$O $J$=1-0 (green contours) intensity map overlayed on $^{12}$CO $J$=1-0 integrated intensity map (color scale). The integrated velocity range is from  19 to 24 km s$^{-1}$.  The contour levels of each CO molecule are 30, 40,..., $90\%$ of the
peak value. Letters A, B, C, and D
indicate the different molecular clumps. The dot symbols mark the mapping
points. The pink triangle represents the IRAS 19282+1814 source. The red and green crosses represent the submillimeter sources and exciting star HD 231616, respectively. The unit of the grey and color bar is in MJy sr$^{-1}$ and K km s$^{-1}$,respectively.}
\end{figure*}

\subsection{H {\small II} region G53.54-0.01}
The concentration of YSOs around H {\small II} region G53.54-0.01 indicates that their formation might have been triggered. The PAH emission of H {\small II} region G53.54-0.01 shows the cometary globule.  To analyze in greater detail the morphology of molecular gas associated with G53.54-0.01, we mapped the emission in the $^{12}$CO $J$=1-0, $^{13}$CO $J$=1-0, and C$^{18}$O $J$=1-0 transitions. Adopting the velocity range of  20 to 28 km s$^{-1}$ from the right panel of Fig. 1, the integrated intensity maps of $^{13}$CO $J$=1-0 and C$^{18}$O $J$=1-0 are made shown in Fig.4 (a). From  Fig.4 (a), we see that the emission of $^{13}$CO $J$=1-0 shows the arc-like structure consistent with the PAH emission at the northwest of G53.54-0.01. Four clumps are identified in C$^{18}$O $J$=1-0, designated clumps A, B, C and D, respectively.
There are two submillimeter-wavelength continuum sources and an IRAS source that are associated with the molecular emission (Di Francesco et al. \cite{Di08} \& Sun et al. \cite{Sun03}). Moreover, JCMTSF J193024.2+182026 and IRAS 19282+1814 is nearly located on the center of clump A. Sun et al. (\cite{Sun03}) suggested that IRAS 19282+1814 may be a massive YSO candidate.

Figure 5 shows the $^{12}$CO $J$=1-0, $^{13}$CO $J$=1-0, and C$^{18}$O $J$=1-0 spectra toward the peak positions of each clump. The line profiles of $^{12}$CO $J$=1-0 and $^{13}$CO $J$=1-0 in clump A appear to only show the blue wings. For clump B, the line profiles of $^{12}$CO $J$=1-0 and $^{13}$CO $J$=1-0 only are broadened in the red wings, while that of C$^{18}$O $J$=1-0 shows double peaks. The line profiles of $^{12}$CO $J$=1-0 and $^{13}$CO $J$=1-0 in clumps C and D is the same as that in clump B, but both line profiles of C$^{18}$O $J$=1-0 show a single peak. The profile of each line is divided into a main and a residual part in the red wing. We fitted each spectrum with two Gaussian profiles. Table 1 shows the fitted results. Adopting two velocity ranges of  19 to 24 km s$^{-1}$ and 24 to 27 km s$^{-1}$, we further made the integrated intensify maps of $^{12}$CO $J$=1-0, $^{13}$CO $J$=1-0 and C$^{18}$O $J$=1-0 shown in Fig.4 (b), (c), and (d). In Fig.4 (b) and (d), the emission maps of $^{12}$CO $J$=1-0, $^{13}$CO $J$=1-0 and C$^{18}$O $J$=1-0  clearly show four clumps. The velocity component of 24 to 27 km s$^{-1}$ is not associated with the clumps A, B, C, and D in Fig.4 (c), but well consistent with the PAH emission of H {\small II} region G53.54-0.01. Hence, we suggest that the velocity component of 24 to 27 km s$^{-1}$  should be another cloud clump, which is weaker than the emission of 19 to 24 km s$^{-1}$, as seen in Table 1.  The velocity component of 24 to 27 km s$^{-1}$  may be collected into the clumps A, B, C, and D with the expansion of  H {\small II} region G53.54-0.01. C$^{18}$O $J$=1-0 is used to trace the dense clump. Comparing the C$^{18}$O $J$=1-0 emission in Fig.4 (b) with that in Fig.4 (c), we found that the position of only clump B has the strong C$^{18}$O $J$=1-0 emission with the velocity component of 24 to 27 km s$^{-1}$, which is responsible for the double-peaks profile of C$^{18}$O $J$=1-0 in the clump B. One of peaks at $\sim$23 km s$^{-1}$ is associated with clump B, another at $\sim$25 km s$^{-1}$ may belong to the collected gas.

Anderson et al. (\cite{Anderson11}) gave that the hydrogen radio recombination line (RRL) velocity is 23.9$\pm$0.3 km s$^{-1}$ for H {\small II} region G53.54-0.01, which are well associated with the main systemic velocities of each clump in C$^{18}$O $J$=1-0 line. According to the Galactic
rotation model of Fich et al. (\cite{Fich89}) together with $R_{\odot}$ = 8.5 kpc and $V_{\odot}$ = 220 km s$^{-1}$, where
$V_{\odot}$ is the circular rotation speed of the Galaxy, we obtain
a kinematic distance of ~2.0 or 8.1 kpc to H {\small II} region G53.54-0.01. Hunter \& Massey (\cite{Hunter90}) gave that H {\small II} region G53.54-0.01 has a  photometric distance of 1.7 kpc.  Hereafter, we will adopt the near kinematic distance of ~2.0 kpc. Using the the optical thin $^{13}$CO $J$=1-0 line and assuming local thermodynamical equilibrium (LTE) we estimate the H$_{2}$ column density towards each clump and the collected gas in Fig. 5. Moreover, we assume that the collected gas has the same excitation temperature as each clump. The column density  are determined by Garden et al. (\cite{Garden})
\begin{equation}
\mathit{N_{\rm ^{13}CO}}=4.6\times10^{13}\frac{(T_{\rm
ex}+0.89)}{\exp(-5.29/T_{\rm ex})}\int T_{\rm mb}\rm dv ~\rm cm^{-2},
\end{equation}\indent
where $\rm dv$ is the velocity range in km s$^{-1}$, and $T_{\rm ex}$ is
the excitation temperature in K.  We estimate  $T_{\rm ex}$ following the
equation
\begin{equation}
\mathit{T_{\rm ex}}=5.53/{\ln[1+5.53/(T_{\rm mb}+0.82)]},
\end{equation}
where $T_{\rm mb}$ is the corrected main-beam temperature of $^{12}$CO
$J$=1-0.  Here we
use the relation $N_{\rm H_{2}}/N_{\rm ^{13}CO}$ $\approx$
$5\times10^{5}$  (Simon et al. \cite{Simon01}) and Gaussian fit line parameters to estimate the H$_{2}$ column density.

\begin{figure*}[]
\vspace{0mm}
\includegraphics[angle=0,scale=0.85]{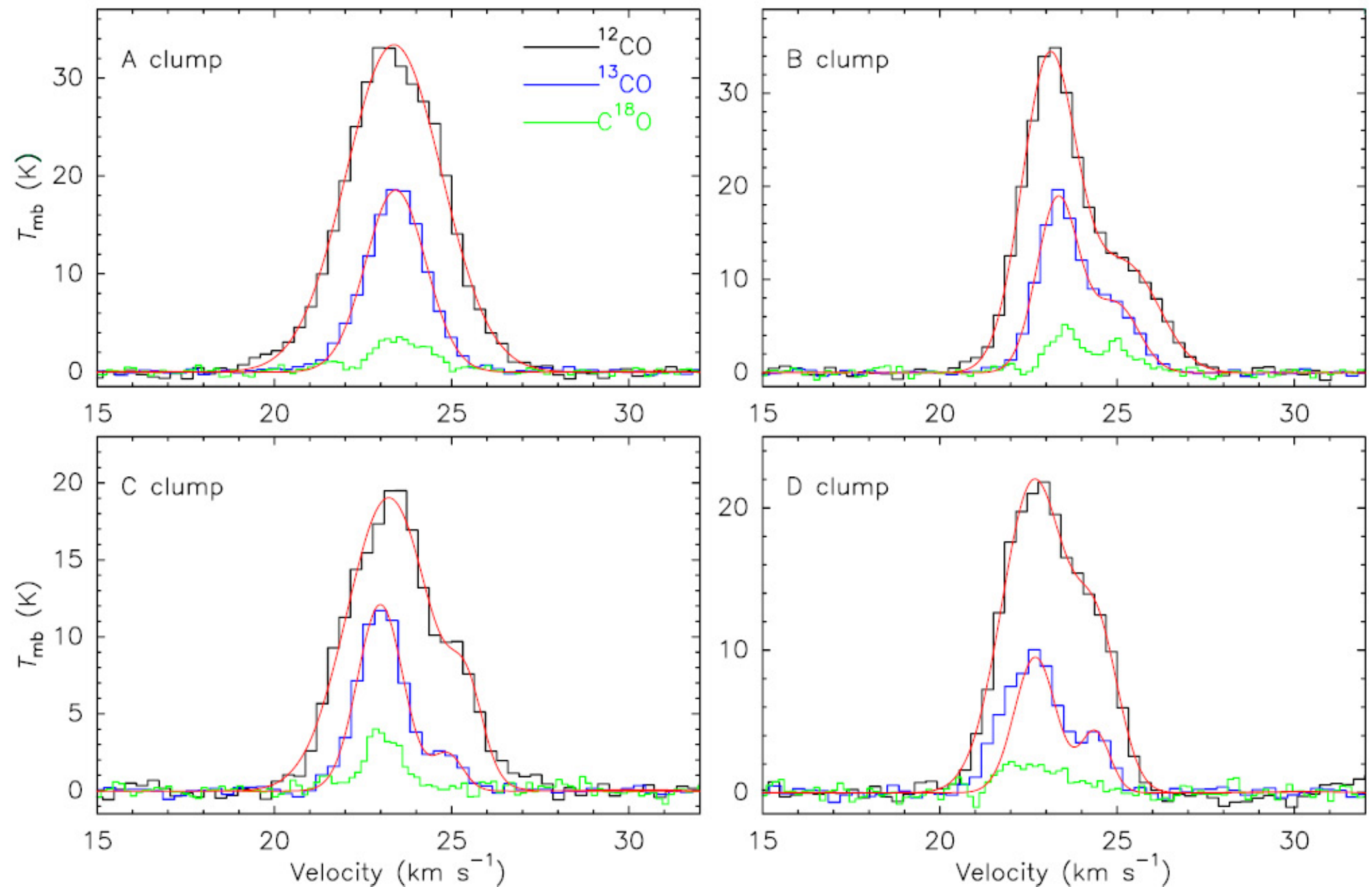}
\vspace{0mm}\caption{$^{12}$CO
$J$=1-0, $^{13}$CO $J$=1-0, and C$^{18}$O $J$=1-0 spectra at the peak position of the molecular cores A, B, C, and D. To better show the profile of C$^{18}$O $J$=1-0 line,
only C$^{18}$O $J$=1-0 intensity is amplified to 1.5 times of the original intensity. The red lines represent the Gaussian fitted lines.}
\end{figure*}

\begin{table*}
\vspace{-2mm}
\begin{center}
\tabcolsep 2.4mm\caption{Observed parameters of each line for each clump}
\def\temptablewidth{10\textwidth}%
\vspace{-2mm}
\begin{tabular}{lccccccccccccc}
\hline\hline\noalign{\smallskip}
Name   &&      & $^{12}$CO J=1-0   &   &&   & $^{13}$CO J=1-0 &  && & C$^{18}$O J=1-0& \\
\cline{3-5}\cline{7-9}\cline{11-13}
        &&   $T_{\rm mb}$   &FWHM  &$V_{\rm LSR}$   &&   $T_{\rm mb}$   &FWHM  &$V_{\rm LSR}$&&   $T_{\rm mb}$   &FWHM  &$V_{\rm LSR}$   \\
        &&  (K)               &(km $\rm s^{-1}$) &(km $\rm s^{-1}$) &&   (K)   &(km $\rm s^{-1}$)  &(km $\rm s^{-1}$)&&   (K)   &(km $\rm s^{-1}$)&(km $\rm s^{-1}$) \\
\hline\noalign{\smallskip}
Clump A &Main     & 33.5 & 3.2 & 23.4  &&  18.6   & 2.0 & 23.4  & &2.2  & 1.9 & 23.6 \\  
Clump B &Main     & 33.8 & 1.8 & 23.1  &&  18.8   & 1.4 & 23.3  & &3.2  & 1.0 & 23.5 \\  
        &Residual & 11.2 & 2.2 & 25.3  &&   7.5   & 1.5 & 25.0  & &2.0  & 1.2 & 25.0 \\  
Clump C &Main    & 19.1 & 2.8 & 23.2  &&  12.1   & 1.5 & 23.0  & &2.6  & 1.3 & 23.1 \\  
        &Residual &  4.8 & 1.1 & 25.5  &&   2.3   & 1.0 & 24.9  & &--   & --  & -- \\  
Clump D &Main     & 22.0 & 2.1 & 22.7  &&   9.6   & 1.9 & 22.6  & &1.4  & 2.2 & 22.6 \\  
        &Residual & 10.2 & 1.5 & 24.5  &&   3.6   & 0.9 & 24.5  & &--   & --  & -- \\  
\noalign{\smallskip}\hline
\end{tabular}\end{center}
\end{table*}

\begin{table*}[]
\vspace{-2mm}
\begin{center}
\tabcolsep 5.8mm \caption{The physical parameters of the clumps in
LTE.}
\def\temptablewidth{10\textwidth}
\begin{tabular}{ccccccccc}
\hline\hline\noalign{\smallskip}
Name   &&$T_{\rm ex}$&$d$&$\int T_{mb}dv$ &$N_{\rm H_{2}}$ & $n(\rm H_{2})$& $M$\\
       && K& (pc)  &(K km $\rm s^{-1}$)  &(cm$^{-2}$) &(cm$^{-3}$)  & ($\rm M_{ \odot}$)  \\
  \hline\noalign{\smallskip}
Clump A  &Main&    37.0&2.1 & 113.5 &1.2$\times10^{23}$  & 9.3$\times10^{3}$ & 3.0$\times10^{3}$  \\  
Clump B  &Main&    37.3&3.2 & 65.7  &6.7$\times10^{22}$  & 3.4$\times10^{3}$ & 3.9$\times10^{3}$ \\  
         &Residual&    &    & 25.6  &2.6$\times10^{22}$  & 1.3$\times10^{3}$ & 1.5$\times10^{3}$ \\  
Clump C  &Main&    22.6&0.9 & 56.5  &3.9$\times10^{22}$  & 7.0$\times10^{3}$ & 1.8$\times10^{2}$ \\  
         &Residual&    &    &  5.6  &3.8$\times10^{21}$  & 0.7$\times10^{3}$ & 0.2$\times10^{2}$   \\  
Clump D  &Main    &25.5&1.2  & 49.9 &3.7$\times10^{22}$  & 5.0$\times10^{3}$ & 3.5$\times10^{2}$ \\  
         &Residual&    &    & 16.2  &1.2$\times10^{22}$  & 1.6$\times10^{3}$ & 1.0$\times10^{2}$  \\  
\noalign{\smallskip}\hline
\end{tabular}\end{center}
\end{table*}

If the clumps are
approximately spherical in shape, the mean number density of $\rm
H_{2}$ is estimated to be
\begin{equation}
\mathit{n(\rm H_{2})}=1.62\times10^{-19}N_{\rm H_{2}}/d,
\end{equation}
where $d$ is the averaged diameter of the clumps in parsecs (pc), measured from Fig. 4.

Moreover, their mass
is given by
\begin{equation}
\mathit{M_{\rm H_{2}}}=\frac{1}{6}\pi
d^{3}\mu_{g}m(\rm H_{2})n(\rm H_{2}),
\end{equation}
where $\mu_{g}$=1.36 is the mean atomic weight of the gas, and $m(\rm
H_{2})$ is the mass of a hydrogen molecule. The obtained results are all listed in Table 2.

\begin{figure}[]
\vspace{0mm}
\includegraphics[angle=0,scale=0.62]{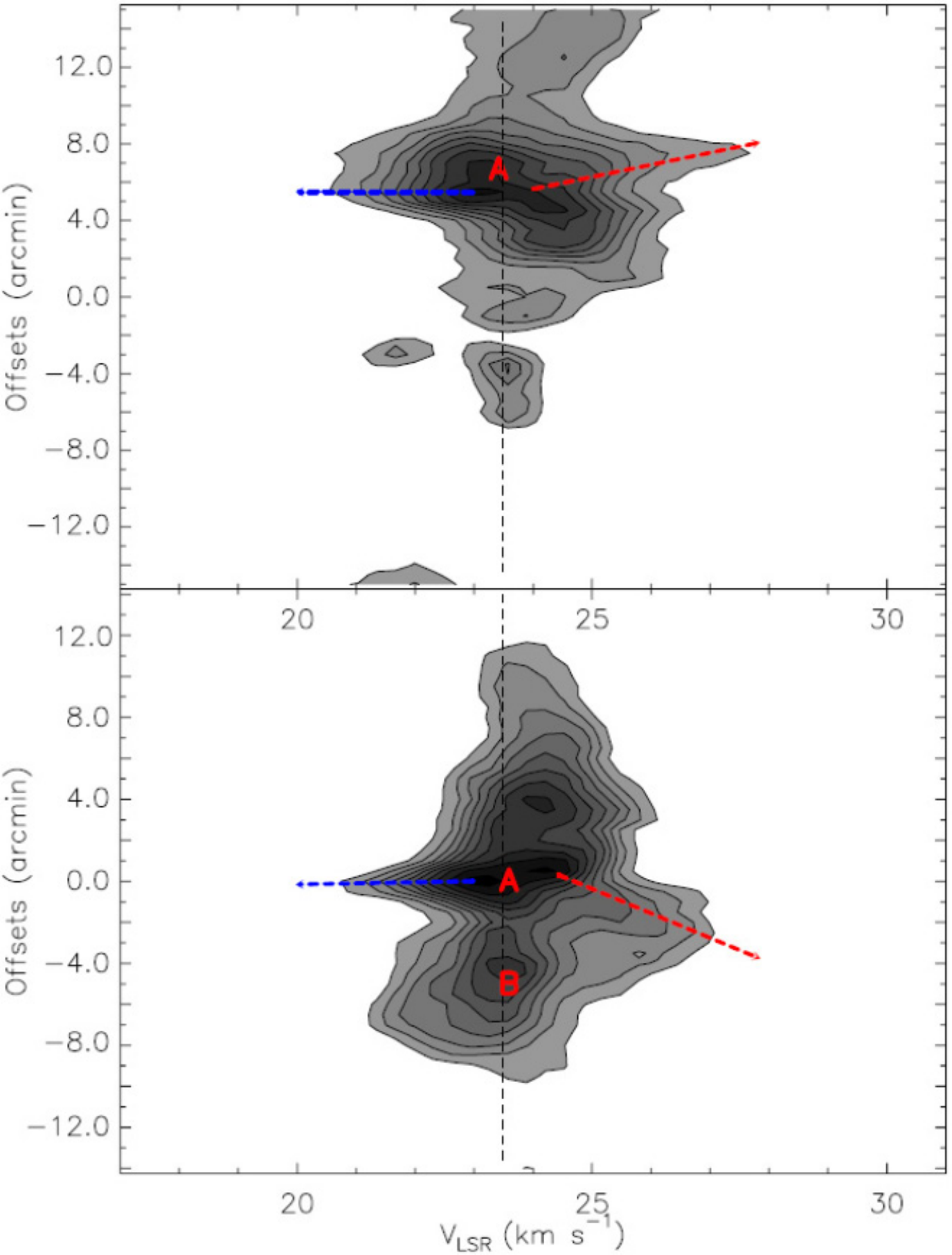}
\vspace{0mm}\caption{P-V diagram constructed from  $^{12}$CO $J=1-0$
transition for clump A. The contour levels are
10, 20,..., $90\%$ of the peak value, with the cut through the peak
position of clump A along the east-west (upper panel) and north-south (under panel) directions. The blue and red dashed arrows show the blueshifted and blueshifted components of
the outflow. The dotted line in the spectra marks
the systemic velocity of clump A. }
\end{figure}

\section{DISCUSSION}
\subsection{The dynamics of the large-scale infrared bubble and H {\small II} region G53.54-0.01}
The large-scale infrared bubble shows the half-shell structure at 8 $\mu$m, which just surrounds a large radio shell centered at $l$=53.9$^{\circ}$ and $b$=0.2$^{\circ}$. Leahy et al (\cite{Leahy08}) suggested that the radio shell may belong to SNR G53.9+0.2 at a distance of $\simeq$7.3 kpc. The large-scale infrared bubble contains three H {\small II} regions (G53.54-0.01, G53.64+0.24, and G54.09-0.06), while a Crab-like young SNR G54.1+0.3 is situated on the radio shell. Because the kinematic distance of SNR G54.1+0.3 is $\simeq$ 6.2 kpc  from HI and CO observations  (Leahy et al. \cite{Leahy08}), SNR G54.1+0.3 is not related to the radio shell. Using the GRS data, we found that the large-scale infrared bubble is associated with several $^{13}$CO $J$=1-0 clumps. The clumps have two velocity components, peak at $\sim$24 and 38 km s$^{-1}$, respectively. The velocity component peak at 38 km s$^{-1}$ coincides with H {\small II} regions G53.64+0.24 and G54.09-0.06, while that at 24 km s$^{-1}$ is associated with H {\small II} region G53.54-0.01. Hence, the 8 $\mu$m emission  consistent with H {\small II} region G53.54-0.01 should belong to the foreground emission, and only overlap with the large-scale infrared bubble in the line of sight. Additionally, the 8 $\mu$m emission of the inner border is more dense in the large-scale infrared bubble. Moreover, the $^{13}$CO $J$=1-0 emission related to H {\small II} region G54.09-0.06 shows a comet-like morphology. We conclude that the large-scale infrared bubble is likely to be formed by ultraviolet (UV) radiation from the progenitor star of SNR 53.9+0.2.

In addition, H {\small II} region G53.54-0.01 is close to a filamentary molecular cloud. We further made the observations in $^{12}$CO $J$=1-0, $^{13}$CO $J$=1-0, and C$^{18}$O $J$=1-0 towards H {\small II} region G53.54-0.01 and its adjacent region. Four clumps are found in this region.  The profiles of $^{12}$CO $J$=1-0 in clump A only show the blue wing. The clump A is associated with IRAS 19282+1814. Sun et al. (\cite{Sun03}) concluded that IRAS 19282+1814 has a blue-shifted monopolar molecular outflow. To confirm the outflow, we made two position-velocity (PV) diagrams (Fig. 6) with the cuts through the
peak position of clump A along the north-south and east-west directions in $^{12}$CO $J$=1-0 line. Clump A shows the bipolar structure presented in the  blue and red dashed arrows.  The blueshifted component has obvious velocity gradients with respect to the systemic velocity of clump A. The redshifted component also has the velocity gradient along the inverse direction, but the gradient is not very obvious relative to the blueshifted component. The
distributions of redshifted and blueshifted velocity components are
an indication of outflow motion.  However, the profiles of $^{12}$CO $J$=1-0 and $^{13}$CO $J$=1-0 in clumps B, C and D only are broadened in red wings. Comparing with the profile of C$^{18}$O $J$=1-0 in clump B, the red wing may not be broadened, but indicate a component. The component may be the collected gas, which expand into the molecular cloud containing clumps A, B, C and D, after collected with the expansion of H {\small II} region G53.54-0.01.

\subsection{Triggered star formation}
Koo et al. (\cite{Koo08}) detected a star-forming loop around SNR G54.1+0.3. They suggested that the star-forming loop may be triggered by the progenitor star of G54.1+0.3. Because of the selected Class I and Class II sources in a large-scale region, we did not find the star-forming loop around SNR G54.1+0.3. Three EGOs sources are found located in the G54.09-0.06 complex, as well as over three H {\small II} regions and two  small-scale bubbles. This distribution is an indication of an active star-forming region located in the large-scale infrared  bubble around SNR G53.9+0.2.

Class I and Class II sources are mostly concentrated around H {\small II} region G53.54-0.01. H {\small II} region G53.54-0.01 is expanding into a filament molecular cloud. Four clumps are found around H {\small II} region G53.54-0.01, but only the clump A associated with a massive YSO (IRAS 19282+1814) may has an outflow motion. Adopting the angle of 90$^{\circ}$ and the clump size as the outflow size, the dynamic timescale of the outflow is estimated roughly by equation t = 9.78$\times$ $10^{5}$R/V (yr), where V in km $\rm s^{-1}$ is the maximum flow velocity relative to the cloud systemic velocity, and R in pc is the outflow size. The average dynamical timescale of the outflow is 4.3$\times$$10^{5}$ yr. The clump A has the mass of 3.0$\times10^{3}$$\rm M_{ \odot}$, indicating a massive star forming region. The above analysis suggest that the triggered star formation have occurred in the region around H {\small II} region G53.54-0.01. The dynamical age of H {\small II} region G53.54-0.01 can also be used to  decide whether YSOs are triggered by H {\small II} region. Assuming the emission is optically thin free-free thermal continuum, the ionizing luminosity $Q_{\rm Ly}$ is computed by Condon (\cite{Condon92})
\begin{equation}
 \mathit{Q_{\rm Ly}}=7.54\times10^{46}(\frac{\nu}{\rm GHz})^{0.1}(\frac{T_{e}}{\rm K})^{-0.45}(\frac{S_{\nu}}{\rm Jy})(\frac{D}{\rm kpc})^{2}\rm ~s^{-1},
\end{equation}\indent
Where $\nu$ is the frequency of the observation, $S_{\nu}$ is the observed specific flux density, and $D$ is the distance to the H {\small II} region.
Anderson et al. (\cite{Anderson11}) gave that HII region G53.54-0.01 has the flux density of 1.3 Jy at 9 GHz. Because the spectral type of the exciting star of H {\small II} region G53.54-0.01 is B0V (Hunter \& Massey \cite{Hunter90}), we adopt an electron temperature of 33340 K (Vacca et al. \cite{Vacca96}). Finally,  we obtain $Q_{\rm Ly}$$\backsimeq$4.5$\times10^{45}$ ph s$^{-1}$.

Using a simple model described by Dyson \& Williams (\cite{Dyson80}) and assuming an H {\small II} region expanding in a homogeneous medium, we estimate the dynamical age of the H {\small II} region as
\begin{equation}
\mathit{t_{\rm  HII}}=7.2\times10^{4}(\frac{R_{\rm H{\small II}}}{\rm pc})^{4/3}(\frac{Q_{\rm Ly}}{10^{49} \rm ph~s^{-1}})^{-1/4}(\frac{n_{\rm i}}{10^{3}\rm cm^{-3}})^{-1/2} \rm ~yr,
\end{equation}\indent
where $R_{\rm HII}$ is the radius of the H {\small II} region, $n_{\rm i}$ is the initial number density of the gas, and $Q_{\rm Ly}$ is the ionizing luminosity.  Adopting the measured radius of H {\small II} region G53.54-0.01 ($\sim$2.3 pc) obtained from Fig. 5, and assuming an initial number density of $\sim$10$^{3}$cm$^{-3}$, we derive a dynamical age of 1.5$\times10^{6}$ yr for H {\small II} region G53.54-0.01.

Comparing the age of H {\small II} region G53.54-0.01 with that of YSOs (Class I and Class II sources) showed on its border, we suggest that these YSOs are likely to be triggered by H {\small II} region G53.54-0.01.  Two processes have been considered for the triggering of star formation at the edge of the H {\small II} regions (Deharveng et al. \cite{Deharveng10}), namely `collect and collapse (CC)' and `radiation
driven implosion (RDI)'. In the CC process, a compressed layer of high-density neutral material forms between the ionization front and shock front preceding it in the neutral gas, and star formation occurs when this layer becomes gravitationally unstable, which is characterized by the presence of the fragments regularly spaced along the molecular ring or shell; In RDI process, the shocks drive into pre-existing density structures and compress them to form stars, which is characterized by the cometary globules or optically bright rims.
The PAH emission of H {\small II} region G53.54-0.01 shows the cometary globule, but we detected the collected gas in the clumps B, C, and D. Furthermore, we identified an outflow only in the clump A associated with the massive YSO IRAS 19282+1814,  but did not detect the collected gas.
We suggest that the collected gas expands into the pre-existing clump A, the local density of clump A
increases then it collapse to be a massive star. The CC process is responsible for the YSOs and an massive star formation around H {\small II} region G53.54-0.01. This process not only collect the diffuse gas near the pre-existing molecular cloud, but also add the collected gas into the pre-existing molecular cloud.

\section{Conclusions}
Using the Spitzer-IRAC 8
$\mu$m and GRS $^{13}$CO $J$=1-0 archival data, we have studied a large-scale infrared bubble centered at $l$=53.9$^{\circ}$ and $b$=0.2$^{\circ}$. H {\small II} regions G53.54-0.01, G53.64+0.24, and G54.09-0.06 are located on the bubble. Molecular observations in $^{12}$CO $J$=1-0, $^{13}$CO $J$=1-0 and C$^{18}$O $J$=1-0 with the Purple Mountain Observation (PMO) 13.7 m radio telescope were performed to investigate the detailed distribution of molecular material associated with HII region G53.54-0.01 (Sh2-82).  The results can be summarized as follows:
 \begin{enumerate}
      \item
 The large-scale infrared bubble presents a half-shell morphology at 8 $\mu$m. H {\small II} regions G54.09-0.06 and G53.64+0.24 have a recombination line velocities of 42.1$\pm$0.8 km s$^{-1}$ and 38$\pm$1.8 km s$^{-1}$, respectively, while 23.9$\pm$0.3 km s$^{-1}$ for H {\small II} region G53.54-0.01,  hence we concluded that the 8 $\mu$m emission related to H {\small II} region G53.54-0.01 should belong to the foreground emission, and only overlap with the large-scale infrared bubble in the line of sight.
      \item
 The large-scale infrared bubble has the dense inner border and several pillar-like structures. Moreover, the $^{13}$CO $J$=1-0 emission related to H {\small II} region G54.09-0.06 shows a comet-like morphology,  suggesting that the large-scale infrared bubble is likely to be formed by ultraviolet (UV) radiation from the progenitor star of old SNR 53.9+0.2.
      \item
Three EGOs sources, three H {\small II} regions and two small-scale bubbles (N116 and N117) are found situated at the G54.09-0.06 complex, suggesting an active massive star-forming region. We suggest that the velocity of  N116 and N117 may be $\sim$38 km s$^{-1}$.
      \item
In $^{18}$CO $J$=1-0 line, we found four cloud clumps on the northeastern border of  H {\small II} region G53.54-0.01. Comparing the spectral profiles of $^{12}$CO $J$=1-0 and $^{13}$CO $J$=1-0  with that of $^{18}$CO $J$=1-0 peak at each clump, we found a collected gas component, which has rushed into three clumps B, C, and D, except for the clump A.
     \item
clump A  with a mass of 3000 $M_{ \odot}$ is associated with a submillimeter-wavelength continuum source and a massive YSO. Additionally, the clump A has an outflow motion. The estimated dynamical timescale of the outflow is 4.3$\times$$10^{5}$ yr, indicate a forming massive star. The derived ages of G53.54-0.01 is 1.5$\times10^{6}$ yr. The significant enhancement of several Class I and Class II YSOs around G53.54-0.01 indicates the presence of some recently formed stars. Taking into account the
age of G53.54-0.01 and YSOs,  we find that G53.54-0.01
may trigger the formation of these YSOs via the collect and collapse (CC) process.
   \end{enumerate}

\begin{acknowledgements}
We are very grateful to the anonymous referee for his/her helpful comments and suggestions.
We are also grateful to the staff at the Qinghai Station of PMO for their assistance during the observations. Thanks for the Key Laboratory
for Radio Astronomy, CAS, for partly supporting the telescope
operation. This work was supported by the National Natural Science Foundation of China (Grant No. 11363004). Also supported by the
young researcher grant of national astronomical observatories,
Chinese academy of sciences.
\end{acknowledgements}

\end{document}